\documentclass{emulateapj}

\newcommand{\text}[1]{\rm #1}

\slugcomment{}

\shorttitle{Stochastic Nature of Gravitational Waves from Supernova Explosions}
\shortauthors{Kotake et al.}

\begin{document}


\title{Stochastic Nature of Gravitational Waves from Supernova Explosions with Standing Accretion Shock Instability}


\author{Kei Kotake\altaffilmark{1}, Wakana Iwakami\altaffilmark{2},
Naofumi Ohnishi\altaffilmark{2}, and Shoichi Yamada\altaffilmark{3,4}}
\altaffiltext{1}{Division of Theoretical Astronomy/Center for Computational Astrophysics, National Astronomical Observatory of Japan, 2-21-1, Osawa, Mitaka, Tokyo, 181-8588, Japan}
\altaffiltext{2}{Department of Aerospace Engineering, Tohoku University,
6-6-01 Aramaki-Aza-Aoba, Aoba-ku, Sendai, 980-8579, Japan}
\altaffiltext{3}{Science \& Engineering, Waseda University, 3-4-1 Okubo, Shinjuku,
Tokyo, 169-8555, Japan}
\altaffiltext{4}{Advanced Research Institute for Science and Engineering, Waseda University, 3-4-1 Okubo, Shinjuku,Tokyo, 169-8555, Japan}

\begin{abstract}
We study properties of gravitational waves based on the  
three-dimensional simulations, which demonstrate the neutrino-driven
explosions aided by the standing accretion shock instability (SASI).
Pushed by evidence supporting slow rotation prior to core-collapse, 
 we focus on the asphericities in neutrino emissions and matter motions outside the 
  protoneutron star. By performing a ray-tracing calculation in 3D,
 we estimate accurately 
 the gravitational waveforms from anisotropic neutrino emissions.
In contrast to the previous work assuming axisymmetry,
we find that the gravitational waveforms vary much more stochastically
because the explosion anisotropies depend sensitively on the growth of the SASI
which develops chaotically in all directions.
Our results show that the gravitational-wave spectrum has its peak near $\sim 100$ Hz, 
reflecting the SASI-induced matter overturns of $\sim O(10)$ ms.
 We point out that the detection of such signals, possibly visible to 
the LIGO-class detectors for a Galactic supernova, could be an important  
probe into the long-veiled explosion mechanism.
\end{abstract}

\clearpage

\section{Introduction}
The gravitational-wave astronomy is now becoming reality.
In fact, significant progress has been made on the
gravitational wave detectors, such as
 LIGO,
VIGRO,
GEO600,
TAMA300, 
and AIGO
with their 
international network of the observatories 
(e.g., \citet{hough} for a review).
For the detectors, core-collapse supernovae
especially in our Galaxy, have been supposed as one of the most plausible sources 
of gravitational waves (see, for example, \citet{kotake_rev,ott_rev} 
for recent reviews). 

Traditionally, most of the model calculations of gravitational waves (GWs)
have focused on the bounce signals
in the context of rotational 
(e.g., \citet{zweg,kotakegw,shibaseki,ott,dimmelprl,simon} and references
therein) and magneotrotational
core-collapse \citep{kotakegwmag,obergaulinger,cerda}.
However recent stellar evolution calculations 
suggest that such rapid rotation assumed in most of the previous studies,
 albeit attracting much attention currently as a relevance 
to collapsar \citep{hirschi,wh},
is not canonical for progenitors of core-collapse supernovae with neutron star formations
\citep{heger05,ott_birth}

In the case of the slowly rotating supernova cores,
two other ingredients are expected to be important in 
the much later phases after core bounce, namely convective motions and 
anisotropic neutrino emissions.  
Thus far, various physical ingredients for producing asphericities and the resulting GWs
 in the postbounce phase, have been studied such as the roles 
of pre-collapse density inhomogeneities \citep{burohey,muyan97,fryersingle},
 moderate rotation of the iron core \citep{mueller}, 
g-mode oscillations of protoneutron stars (PNSs) 
\citep{ott_new},  and the standing-accretion-shock-instability (SASI)
 \citep{kotake_gw_sasi,marek_gw}.

However, most of them have been based on two-dimensional (2D) 
simulations that assume axisymmetry. 
  Then, the growth of SASI (e.g., \citet{blondin_03,scheck_04,ohnishi_1,fogli07}) 
and the large-scale convection,  
both of which are now considered
to generically develop in the postbounce phase 
and to help the neutrino-driven explosions \citep{bethe,mj}, develop
along the symmetry axis preferentially, thus suppressing the
anisotropies in explosions.
So far very few three-dimensional (3D) studies have
been conducted \citep{muyan97,fryer04}. Moreover, 
neither the growth of the SASI  
nor its effects on the GWs have been studied yet
because SASI is suppressed artificially owing to the 
limited computational domain \citep{muyan97} or to the early
shock-revival \citep{fryer04}, which has not been discovered by other
supernova simulations.

In this {\it Letter}, we study the properties of the gravitational radiation 
 based on the 3D simulations, which demonstrate the neutrino-driven
explosions aided by SASI \citep{iwakami08}. 
Supported by the evidence of the slow rotation prior to core-collapse \citep{heger05,
ott_birth},
 we focus on the asphericities outside the protoneutron stars,
 which are produced by the growth of SASI. By performing a ray-tracing calculation in 3D 
\citep{kotake_ray}, 
we estimate accurately the gravitational waveforms from anisotropic neutrino emissions.
 We show that the features of the gravitational waveforms
are significantly different than the ones in the axisymmetric cases, 
which should tell us the necessity of the 3D supernova modeling.


\section{Numerical Methods and Models}\label{sec2}
The employed numerical methods and model concepts are essentially the same as 
those in our previous paper \citep{iwakami08}. 
Using the ZEUS-MP \citep{hayes} as a hydro-solver, we solve the 
dynamics of the standing accretion shock flows of matter 
attracted by the protoneutron star and irradiated by neutrinos 
emitted from the protoneuton star.
We employ the so-called light-bulb approximation (see, e.g.,
 \citet{jankamueller96}) and adjust the neutrino luminosities from the
PNSs to trigger explosions.

The computational grid is comprised of 300 logarithmically spaced,
radial zones to cover from the absorbing inner boundary of 
$\sim 50 ~{\rm km}$ to the outer boundary of $2000 ~{\rm
km}$,
 and 30 polar ($\theta$) and 60
azimuthal ($\phi$) uniform mesh points, which are used to cover the whole
solid angle (see for the resolution tests in \citet{iwakami08}).

The initial conditions are provided in the same manner of 
\cite{ohnishi_1}, which describes the spherically symmetric steady accretion
flow through a standing shock wave \citep{yamasaki}. To induce
non-spherical instability we add random velocity perturbations to be
less than 1 $\%$ of  the unperturbed velocity.
Changing the input (electron-)neutrino luminosity at the surface of the protoneutron star
from $L_\nu = 6.4 - 6.85 \times 10^{52}$~erg~s$^{-1}$, 
we have computed six 3D models and one 2D model (see Table \ref{table1}.) 
Except for model E, we can observe the continuous increase of the 
average shock radius with the growth of SASI, reaching the outer 
boundary of the computational domain with the
explosion energy of $\sim 10^{51}$ erg. Until this moment, we run the 
simulations (e.g.,  $\Delta t$ in the table).

Following the method in \citet{epstein,muyan97}, 
the two modes of the GWs from anisotropic neutrino emissions are 
 derived as follows,
\begin{eqnarray}
    h_{+} &=& 
C \int_{0}^{t} \int_{4 \pi} d\Omega' 
(1 + s(\theta') c (\phi') s(\xi) + c(\theta') c(\xi)) \times \nonumber \\ 
& & \frac{( s(\theta')c(\phi') c(\xi) - c(\theta')s(\xi))^2 - s^2(\theta') s^2(\phi')}
{[s(\theta')c(\phi')c(\xi) - c(\theta')s(\xi)]^2 + s^2(\theta')s^2(\phi')} 
\frac{dl_{\nu}(\Omega',t')}{d\Omega'},\nonumber
\label{t+}
\end{eqnarray}
and 
\begin{eqnarray}
     h_{\times} &=& 2C  \int_{0}^{t}
\int_{4\pi} d\Omega' (1 + s(\theta') c (\phi') s(\xi) + c(\theta') c(\xi)) \times \nonumber \\ 
& &  \frac{s(\theta') s(\phi')(s(\theta') c(\phi') c(\xi) - c(\theta')s(\xi))}{[s(\theta')c(\phi')c(\xi) - c(\theta')s(\xi)]^2 + s^2(\theta')s^2(\phi')}
 \frac{dl_{\nu}(\Omega',t')}{d\Omega'},\nonumber
\label{tcc}
\end{eqnarray}
where $s(A)\equiv \sin(A), c(B)\equiv \cos B$,
 $C \equiv 2G/(c^4 R)$ with $G,c$ and $R$, being the gravitational constant, 
 the speed of light, the distance of the source to the observer respectively,
 $dl_{\nu}/d\Omega$
represents the direction-dependent neutrino luminosity emitted per unit
of solid angle into direction of ${\bf \Omega}$, and  $\xi$ is the viewing angle 
(e.g., \citet{kotake_ray}).
For simplicity, we consider here two cases, in which the observer is 
 assumed to be situated along 
'polar' ($\xi=0$) or 'equatorial'($\xi=\pi/2$) direction.
To determine 
$dl_{\nu}/d\Omega$, we perform a ray-tracing calculation. 
Since the regions outside the PNSs are basically thin to neutrinos,  
 we solve the neutrino transport equations in a post-processing manner to estimate
 the neutrino anisotropy,
in which neutrino absorptions and emissions by free nucleons, dominant outside the 
PNSs, are taken into account
 (see \citet{kotake_ray} for more details).
To get the numerical convergence, we need to set 45000 rays to determine the luminosity 
 for each given direction of $\bf{\Omega}$. The GW amplitudes from 
mass motions are extracted by using the standard quadrupole formula 
(e.g., \citet{muyan97}),
 and their spectra by \citet{flanagan} with the FFT techniques. 



\begin{deluxetable}{ccccccc}
\tabletypesize{\scriptsize}
\tablecaption{Model Summary \label{table1}}
\tablewidth{0pt}
\tablehead{\colhead{Model}
 & \colhead{$L_{\nu}$ }
 & \colhead{$\Delta t$}
 & \colhead{$|h^{\rm p}_{\rm max}|$} 
 & \colhead{$|h^{\rm e}_{\rm max}|$} 
 & \colhead{$E_{{\rm GW}}$} 
\\
 & \colhead{($10^{52}$ erg/s)}         
 & \colhead{({\rm ms})}
 & \colhead{($10^{-22}$)}
 & \colhead{($10^{-22}$)}
 & \colhead{($10^{-11}M_{\odot}c^2$)}}
\startdata
A    &6.8     & 507    & 3.0($\times$)  &3.1($\times$)  & 6.0 \\ 
B    &6.766   & 512    & 2.4(+)         &3.1($\times$)  & 7.4 \\ 
C    &6.7     & 532    & 3.7($\times$)  & 2.6($\times$  & 7.4 \\ 
D    &6.6     & 667    & 2.0($\times$)  & 2.0(+)        & 10.8 \\ 
E    &6.4     & ---    & 0.7(+)         & 1.6 ($\times$)& 2.6\\ 
F    &6.85    & 429    & 2.1($\times$)  & 4.3(+)        & 5.6\\ 
${\rm A}'$(2D) &6.8    & 509   & --- & 7.7(+)&  4.4   \\         %
\tablecomments{$L_{\nu}$ denotes the input luminosity.  
$\Delta t$ represents the simulation time. Model A' is the 2D
version of model A. 
 $|h^{\rm p}_{\rm max}|$ and $|h^{\rm
e}_{\rm max}|$ represents the maximum amplitudes seen from the pole
or the equator, respectively. $+$ or $\times$ indicates the
polarization of the GWs at the maximum. $E_{{\rm GW}}$ is the total radiated GW energy. The supernova is assumed to be located at
a distance of 10 kpc.}
\end{deluxetable}

\begin{figure}[hbt]
\epsscale{1.0}
\plotone{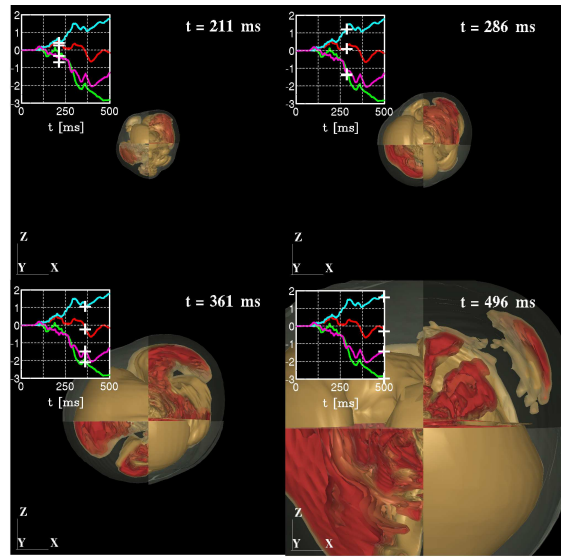}
\caption{Four snapshots of the entropy distributions of model A.
The second and fourth quadrant of each panel shows 
the surface of the standing shock wave.
In the first and third
quadrant, the profiles of the 
high entropy bubbles (colored by red) inside the section cut by the 
$ZX$ plane are shown. The side
length of each plot is 1000km.
The insets show the gravitational waveforms from anisotropic neutrino emissions, 
with '$+$' on each curves 
representing the time of the snapshot. Note that the colors of the curves are
taken to be the same as the top panel of Figure \ref{fig2}.}
\label{fig1}
\end{figure}

\section{Results}\label{sec3}
Figure \ref{fig1} shows the 3D hydrodynamics features of SASI 
from the early phase of the non-linear regime of SASI (top left) until the
shock break-out (bottom right) with the gravitational waveform from neutrinos 
inserted in each panel.

\begin{figure}[hbtp]
\epsscale{1.1}
\plotone{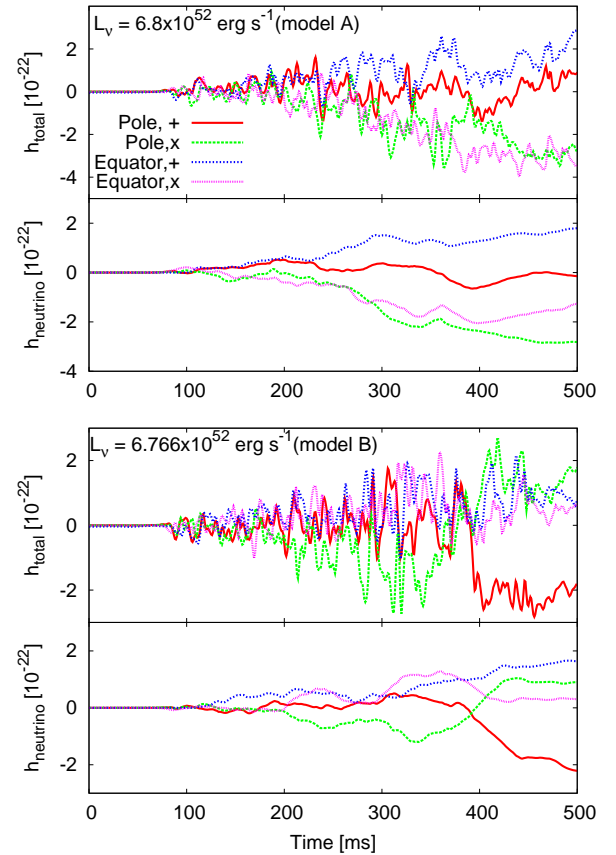}
\caption{Gravitational waveforms from neutrinos (bottom) and from the sum of 
 neutrinos and matter motions (top), seen from the polar
axis and along the equator (indicated by 'Pole' and 'Equator')
with polarization ($+$ or $\times$ modes) for models A and B.
The distance to the SN is assumed to be 10 kpc.}
\label{fig2}
\end{figure}

After about $100$ ms, the deformation of the standing 
shock becomes remarkable marking the epoch when the SASI enters the non-linear regime 
(top left of Figure \ref{fig1}).
At the same time, the gravitational amplitudes 
begin to deviate from zero. Comparing the top two panels in Figure \ref{fig2}, 
which shows the total amplitudes (top) and the neutrino contribution only (bottom),
 it can be seen that the gross structures of the waveforms are 
 predominantly determined by the neutrino-originated GWs with the slower temporal 
 variations ($\gtrsim 30 - 50$ ms), to which the GWs from matter motions 
 with rapid temporal variations ($\lesssim 10$ ms) are superimposed. 
So, we first pay attention to the neutrino GWs in the following and discuss the 
importance of the matter GWs later in the spectrum analysis.

As seen from the top right through bottom left to right panels of Figure
\ref{fig1}, the major axis of the growth of SASI 
is shown to be not aligned with the symmetric axis (:$Z$ axis in the figure)
and the flow inside the standing shock wave is not symmetric with
respect to this major axis (see the first and third quadrant
in Figure \ref{fig1}). This is a generic feature in the computed 3D
models, which is in contrast to the axisymmetric case.
The GW amplitudes from SASI in 2D showed an increasing trend with time  
due to the symmetry axis, along which SASI can develop
preferentially \citep{kotake_gw_sasi,kotake_ray}.
Free from such a restriction, 
 a variety of the waveforms is shown to appear (see waveforms inserted in Figure 1).
Furthermore, the 3D standing shock can also oscillate in all
 directions, which leads to the smaller explosion anisotropy than 2D.
 With these two factors, the maximum amplitudes seen either from the
equator or the pole becomes smaller than 2D (compare 
$|h^{e}_{\rm max}|$ in Table \ref{table1} for models A and A').
On the other hand, their sum in terms of the total radiated
energy ($E_{\rm GW}$ in Table \ref{table1}) are found to be 
almost comparable between 2D and 3D models, which is likely to imply the
energy equipartition with respect to the spatial dimensions.
It was reported in a pioneering 3D study \citep{muyan97} that the emitted GW energy in the forms of neutrinos was reported to
be about 2 orders of magnitude smaller for 3D than in 2D. 
However this may be because the global anisotropy of the SASI
was not captured in
their simulations with the limited computation domain in the polar direction.

Figure \ref{fig2} shows the gravitational waveforms for different
luminosity models. The input luminosity for the two pair panels differs
only $0.5\%$. Despite the slight difference, the waveforms of each polarization 
are shown to exhibit no systematic similarity when seen from the pole or equator.
This is due to the chaotic nature of SASI influenced by small
 differences. No systematic dependence of the maximum amplitudes
 and the radiated GW energies on the neutrino luminosities is 
found among the models investigated here (see Table 1). In fact, the largest emitted 
 GW energy is obtained for model D with the intermediate neutrino
luminosity (Table 1). 
It is noted that the waveforms are even sensitive to the difference of the grid 
zoning due to the stochastic nature, although no significant difference in the 
 GW amplitudes is seen in the numerical run with twice as many grid
 points in the $\theta$ and $\phi$ directions.

\begin{figure}
\epsscale{1.0}
\plotone{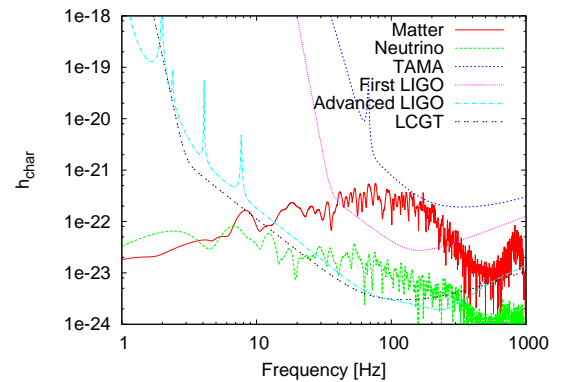}
\caption{Characteristic gravitational-wave spectra from anisotropic neutrino
emissions ('Neutrino') and matter motions ('Matter') with optimal 
detection limits of TAMA, the first LIGO, advanced LIGO, and LCGT for a
supernova at a distance of 10 kpc.
}
\label{fig3}
\end{figure}

Now we move on to discuss the features of the waveforms by the
spectrum analysis. 
From Figure \ref{fig3}, it can be seen that 
the neutrino GWs, albeit dominant over the matter GWs 
in the lower frequencies below $\sim 10$ Hz, 
 become very difficult to detect for ground-based detectors whose sensitivity is 
limited mainly by the seismic noises at such lower frequencies 
\citep{tamanew,firstligonew,advancedligo,lcgt}. 
On the other hand, the gravitational-wave spectra from matter motions peak 
 near  $\sim 100$ Hz, reflecting
 the growth of $\ell = 2$ mode of SASI with timescales of $O(10)$ ms.
Such signals from a Galactic supernova are probably within the detection limits 
of the LIGO-class detectors, and seem surely visible to the next-generation detectors such as the advanced LIGO and LCGT.
 It is noted that another peak in the GW spectra near $\sim$ 1kHz is from 
 the rapidly varying matter motions of $O({\rm ms})$, induced by the local 
hydrodynamical instabilities. 
  These gross properties in the GW spectra are found to be common 
to the other luminosity models. Thus the peak in the spectra near $\sim 100 $Hz 
 is a characteristic feature obtained in the 3D models computed here.

 \section{Summary and Discussion }\label{sec4}
We studied the properties of GWs based on the 3D
simulations, which demonstrate the neutrino-driven explosions with SASI. 
 We focused on the asphericities outside the protoneutron stars,
 which are produced by the growth of SASI.
 By performing a ray-tracing calculation in 3D,
 we estimated accurately the gravitational waveforms from anisotropic 
neutrino emissions. Our results showed that
the waveforms vary much more stochastically than for 2D 
because the explosion anisotropies depend sensitively on the growth of the SASI
which develops chaotically in all directions.
From the spectra analysis, it was pointed out that 
the gravitational-wave spectra from matter motions peak 
 near  $\sim 100$ Hz, reflecting the growth of $\ell = 2$ mode of SASI, 
and that such signals are probably 
visible to LIGO-class 
detectors for a Galactic source.

Here it should be noted that the simulations highlighted in this paper  
are only a very first step towards 
realistic 3D modeling of the supernova explosions.
The approximations adopted in this paper, such as the replacement of
the PNS by the fixed inner boundary and the light-bulb approach with the
constant neutrino luminosity, should be improved.
Important ingredients for the GW emissions such as 
the oscillations of the PNSs \citep{burr_new} and
 the enhanced neutrino emissions near the equators deep inside the PNS observed
 in \citet{marek_gw}, cannot be treated in principle here. 
The exploration of these phenomena will require consistent 
 simulations covering the entire stellar core and 
starting from gravitational collapse 
with neutrino transfer coupled to the 3D magnetohydrodynamics. 
Besides, the effects of rotation 
 on the SASI are yet to be investigated \citep{iwakami08_2}, although the construction
 of the rotating steady accretion flows is still a major undertaking.


Bearing these caveats in mind, 
the stochastic nature of the GWs, which is illuminated here by the 3D models, 
is qualitatively new.
It should be also mentioned that the GW spectra for the acoustic mechanism 
 showed a very strong peak near $\sim$ kHz, reflecting the g-modes
oscillations of the PNS \citep{ott_new}, and are
significantly different from the waveforms discussed here for the 
SASI-aided neutrino-driven models. This means that the successful 
detection of the GW signals will supply us a powerful tool to probe 
the explosion mechanism. 
Recently the next-generation detectors, by which the GWs studied here
could be surely visible for a Galactic source,
are planned to be on line soon 
(2014!) \citep{lantz}. 
Here it is an urgent task for theorists to make precise
 predictions of the GW signals based on more sophisticated 3D supernova modeling.

KK would like to thank K. Sato for continuing 
encouragements.  
KK and SY are grateful to H.-Th. Janka and E. M\"{u}ller 
for their kind hospitality during their stay in MPA.
Numerical computations were carried on XT4 
at CfCA of the NAOJ.  This
study was supported in part by the Grants-in-Aid for the Scientific Research 
from the Ministry of Education, Science and Culture of Japan (Nos. 19540309 and 20740150).

\end{document}